\documentclass[preprint,authoryear,12pt]{elsarticle}

\usepackage{geometry}                
\usepackage{graphicx}
\usepackage{amssymb}
\usepackage{epstopdf}
\usepackage{natbib}
\usepackage{color}
\usepackage[normalem]{ulem}
\usepackage{wasysym}  

\usepackage{lineno}

\journal{Earth Planetary Science Letters}

\begin{document}

\begin{frontmatter}

\title{The Tethered Moon}

\author{Kevin J.\ Zahnle\corref{cor1}\fnref{label1}}
\ead{Kevin.J.Zahnle@NASA.gov}

\cortext[cor1]{Corresponding author}
\address[label1]{Space Science Division, NASA Ames Research Center, MS 245-3, Moffett Field CA 94035 USA}

\author{Roxana Lupu\fnref{label2}}
\ead{roxana.s.lupu@nasa.gov}

\address[label2]{SETI Institute, Mountain View CA 95064, USA}

\author{Anthony Dobrovolskis\fnref{label2}}
\ead{anthony.r.dobrovolskis@nasa.gov}

\author{Norman H.\ Sleep\fnref{label3}}
\ead{norm@stanford.edu}

\address[label3]{Department of Geophysics, Stanford University, Stanford CA 94305, USA}

\begin{abstract}

We address the thermal history of the Earth after the Moon-forming impact,
 taking tidal heating and thermal blanketing by the atmosphere into account.
The atmosphere sets an upper bound of $\sim 100$ W/m$^2$ on how quickly the Earth can cool. 
The liquid magma ocean cools over 2-10 Myrs, with longer times corresponding to high angular-momentum events.
Tidal heating is focused mostly in mantle materials that are just beginning to freeze.
The atmosphere's control over cooling 
sets up a negative feedback between viscosity-dependent tidal heating and temperature-dependent viscosity of the magma ocean. 
While the feedback holds, evolution of the Moon's orbit is limited by the modest radiative cooling rate of Earth's atmosphere. 
Orbital evolution is orders of magnitude slower than in conventional constant $Q$ models, which 
promotes capture by resonances.
 The evection resonance is encountered early, when the Earth is molten.
Capture by the evection resonance appears certain but unlikely to generate much eccentricity 
because it is encountered early when the Earth is molten and $Q_{\oplus}\gg Q_{\footnotesize\leftmoon}$.
Tidal dissipation in the Earth becomes more efficient ($Q_{\oplus}\ll Q_{\footnotesize\leftmoon}$)
later when the Moon is between $\sim\!20R_{\oplus}$ and $\sim\!40R_{\oplus}$.
If lunar eccentricity grew great, this was when it did so, perhaps setting the table for some other process
to leave its mark on the inclination of the Moon.

\end{abstract}


\end{frontmatter}

\noindent Highlights
\begin{itemize}
\item We address the thermal history of the Earth after the Moon-forming impact.
\item Thermal blanketing by Earth's atmosphere limits cooling to $\sim 100$ W/m$^2$.
\item Feedback between tidal heating and viscosity tethers the Moon to Earth.
\item Slow orbital evolution promotes capture by resonances.
\item Tidal Q's evolve differently in Earth and Moon
\end{itemize}

\newpage


\section{Introduction}

The leading explanation for the origin of the Earth-Moon system is that it formed 
when two planets collided 
\citep{Hartmann1986,Benz1986,Stevenson1987,Canup2000,Canup2004,Reufer2012,Cuk2012,Canup2012}.
The event is widely known as the Moon-forming impact.
But it might have been called the Earth-forming impact.
Indeed, in some scenarios \citep[e.g.,][]{Canup2012}, neither of the two colliding planets is markedly more a proto-Earth
than the other, in which case the Moon-forming impact is literally the Earth-forming impact as well.

Here we accept as an initial condition that a Moon-forming impact took place. 
We focus on the angular-momentum-conserving canonical model \citep{Canup2004},
although we also consider
some aspects of recent non-angular-momentum-conserving alternatives \citep{Reufer2012,Cuk2012,Canup2012}
that may better address isotopic evidence that the Earth and Moon share the
same source \citep[e.g.,][]{Pahlevan2007,Zhang2012}.

Our subject here is the earliest thermal evolution of the Earth after the impact. 
Earth's cooling was greatly slowed by the thermal blanketing effect of its atmosphere, which was thick and deep and very opaque,
and Earth's interior was strongly heated by tidal flexing, because the Moon was close by and the Earth was spinning rapidly.
We stress here that tidal heating and thermal blanketing were linked, because radiative cooling by the atmosphere
 limits how big tidal heating can be.
Hence we are also concerned here with the early tidal evolution of the Moon's orbit,
which we will treat as simply as possible.
We first describe the thermal model of the cooling Earth in Section \ref{Earth}.
We then describe tidal heating and tidal evolution in Section \ref{Tides}.  

\section{The cooling Earth}
\label{Earth}
\subsection{Initial conditions}

A reasonable initial condition on the Earth after the Moon-forming impact is that it behaved like a hot liquid 
\citep{Zahnle2006,Elkins-Tanton2008}.  
The impact itself melted and partially vaporized the side of the Earth that was directly struck.
But even if half the mantle survived the shock of collision unmelted \citep{Tonks1993}, it would have collapsed under its own weight
on a Rayleigh-Taylor instability timescale of the order of
\begin{equation}
\tau \approx {\rho \nu \over g \,\Delta\rho\,L},
\end{equation}
where $L$ is the relevant length scale, $g$ gravity, $\Delta\rho/\rho$ the density contrast between the solid and the melt,
and $\nu$ the kinematic viscosity of the solid. 
For reasonable choices ($\Delta\rho/\rho = 0.02$, $L=3\times 10^8$ cm, $\nu = 10^{18}$ cm$^2$/s, the latter appropriate
to a warm silicate planet during accretion),
the solid hemisphere would collapse in just 5 years and in the process releasing     
more than enough gravitational energy to melt what had not yet melted \citep{Tonks1993}.
Other factors --- the impactor's shock-heated molten iron plunging through the mantle to join with Earth's core and the superheated state of the upper core immediately thereafter --- also suggest that a liquid planet is a plausible initial condition.

 It is important here to be careful defining what is meant by melting.
Melting of a silicate is characterized by three temperatures:
 a liquidus $T_{\rm liq}$ hotter than which the silicate is fully molten;
 a solidus $T_{\rm sol}$ colder than which the silicate is fully solid;
 and a critical temperature $T_{\rm crit}$ in between at which the rheology of the material changes from that of a solid
 with melt percolating through it ($T<T_{\rm crit}$) to a liquid carrying suspended solids \citep{Abe1997,Solomatov2009}.
The critical temperature $T_{\rm crit}$ marks a phase transition in viscosity, which 
best corresponds to what is colloquially meant by a melting point.
When the mantle collapses under its own weight, it is viscosity that converts that motion into heat.
Viscous dissipation can only raise the temperature of the mantle to $T_{\rm crit}$.
To raise the temperature of such a lower mantle to the liquidus requires another mechanism,
perhaps a prompt effect like the impactor's shock-heated molten iron dispersing in the mantle before it joins with Earth's core, or 
perhaps a delayed effect stemming from the mantle being strongly heated from below by the core. 

Although very important to geochemistry, for our purposes it is not very important whether the mantle be fully or only partially molten, provided only that initially it behave like a low viscosity liquid.
We therefore begin our study with a liquid mantle and a surface temperature of 3000-4000 K
some 100-1000 years after the impact, by which point we can hope that 
early transients have settled down.
On the other hand, these transients may last long enough to influence moon formation, which is also generally regarded
as very fast \citep{Canup2004,Salmon2012}.

A second initial condition is the presence of a substantial atmosphere.  
Earth today has about 270 bars of H$_2$O and 50 bars of CO$_2$ near the surface, the former mostly in oceans and the latter
mostly in carbonate rocks.  
Similarly large reservoirs are held in the mantle \citep{Sleep2001}.
Thus it is reasonable to start with 100-1000 bars of H$_2$O and CO$_2$ in the atmosphere.
The water vapor and carbon dioxide would be supplemented by smaller amounts of CO, H$_2$, N$_2$, various sulfur-containing gases, and by a suite of geochemical volatiles evaporated from the magma \citep{Schaefer2012,Fegley2013}.

A third initial condition is the existence of the Moon. 
We start the Moon with its current mass and density at the relevant Roche limit of $2.9 R_{\oplus}$.
Although accretion theory suggests that the Moon could accumulate in less than a year,
the need to radiate the energy liberated by lunar accretion likely prolonged the process by decades or even centuries\citep{Thompson1988,Ida1997,Genda2000,Canup2004,Salmon2012}.
During this time interactions between the accreting Moon and the unaccreted debris disk may raise the starting
point of the Moon beyond the Roche limit to perhaps $\sim 5 R_{\oplus}$ \citep{Salmon2012}.
For our purposes starting the Moon nearer the Earth is more interesting because there is more tidal heating and more resonances
to encounter as the Moon's orbit evolves.

A fourth initial condition is the initial angular momentum of the Earth-Moon system.
In canonical models this is held constant, so that with the Moon forming at $2.9 R_{\oplus}$, the Earth
starts with a 5.0-hour day.  Some recent Moon-forming models posit loss
of angular momentum from the Earth-Moon system \citep{Cuk2012,Canup2012}.  
These begin the Earth with a $\sim\!3$-hour day, and thus with significantly more rotational energy available 
to be dissipated as tidal heating.

\subsection{Interior Cooling} 

The total thermal energy $E$ of the Earth is the sum of the thermal energy in the interior $E_{i}$ and the thermal energy in the atmosphere $E_{\rm atm}$.  
Our interest is almost entirely in the time rate of change of $E$,
\begin{equation}
\label{Radiative_cooling}
{\dot E} = -4\pi R_{\oplus}^2\sigma T_{\rm eff}^4 + \pi R_{\oplus}^2F_{\odot}\left(1-A\right) + {\dot E}_{\rm tide} .
\end{equation}
Equation \ref{Radiative_cooling} describes the Earth's cooling rate as the difference between thermal infrared radiation emitted to space by
the atmosphere and the sum of net insolation (radiative heating), tidal heating, and heating from the decay of radioactive elements.
Thermal radiation (radiative cooling) to space is described by an effective radiating temperature $T_{\rm eff}$,
 which is determined by the thermal blanketing effect of the atmosphere and the surface temperature $T_{s}$;
$\sigma$ is the Stefan-Boltzmann constant and $R_{\oplus}$ is the radius of the Earth.
The relation between $T_{\rm eff}$ and $T_s$ --- the greenhouse effect --- is discussed in the next section.
Radiative heating is described by the solar constant $F_{\odot}\approx 1000$ W m$^{-2}$ ca.\ 4.4 Ga
and an albedo $A$. We leave $A=0.3$ as today. 
Tidal heating ${\dot E}_{\rm tide}$ is a big term addressed in Section~\ref{Tides} below. 
With radioactive heating ignored,
the geothermal heat flow $F$ is equivalent to the rate of cooling of the interior,
 \begin{equation}
 4\pi R_{\oplus}^2 F={\dot E}_{\rm tide}-{\dot E_{i}}.
\end{equation}

\subsection{Radiative Cooling} 
\label{Radiative Cooling}

It is well known that an atmosphere's thermal blanketing effect prevents a magma ocean\footnote{The term ``magma ocean'' is broadly used in the literature to describe a mantle with significant partial melt.} from cooling rapidly.
A string of increasingly sophisticated models, beginning with \citet{Matsui1986}
and proceeding through \citet{Abe1988,Zahnle1988}, \citet{Abe1997}, \citet{Zahnle2006}, \citet{Elkins-Tanton2008,Elkins-Tanton2012}, 
\citet{Hamano2013}, and \citet{Lebrun2013} have addressed some of
the consequences of thermal blanketing by steam atmospheres over the Earth's cooling magma oceans.
Much of the progress made in these studies has been in constraining how the magma ocean freezes, how long it takes to freeze, and 
how, when, and what it degasses. 
All of these models presume H$_2$O-CO$_2$ atmospheres, and all but \citet{Zahnle2006} ignore the Moon.

A different set of studies has addressed the composition of the atmosphere
\citep{Schaefer2010,Schaefer2012,Fegley2013,Lupu2014}.
Shortly after the impact the atmosphere over the lava is very hot and contains a great many geochemical volatiles that can evaporate from a magma ocean, such as sulfur, sodium, and chlorine, and many other compounds of H, C, N, and O
in addition to H$_2$O and CO$_2$ \citep{Schaefer2012,Fegley2013}.
The geochemically enriched atmosphere 
 is more opaque than just H$_2$O and CO$_2$ at magma temperatures and thus the planet cools more slowly and spends more time with a lava surface than previous models would predict.

To address these issues we make use of new 1-D non-gray radiative-convective atmospheric structure models 
developed to compute the luminosity and spectral characteristics of Earth-like planets after giant impacts \citep{Lupu2014}.
As the \citet{Lupu2014} models are discussed fully elsewhere we will not provide full documentation here, but we do note
some key features.
The new models include many of the molecular and atomic opacity sources that would be abundant in equilibrium 
with a hot rock or magma surface with the composition of the bulk silicate Earth \citep[BSE,][]{Lupu2014}.
The thermal structure models are computed as equilibrium models in which thermal radiation emitted by the atmosphere
is balanced by the sum of sunlight absorbed and geothermal heat extracted from the cooling Earth.  This assumption
is very good because the heat capacity of the atmosphere is very small compared to the heat capacity of the planet.
A more questionable assumption of the model is that it keeps the atmosphere in chemical equilibrium with the bulk silicate Earth
at any temperature --- i.e., it assumes that CH$_4$ and NH$_3$ form when equilibrium chemistry would favor them.
The assumption is reasonably good at high temperatures (at which CH$_4$ and NH$_3$ are rare in any case),
but for temperatures much below 1000 K equilibrium would not be expected unless effective catalysts
were present among the cloud particles.  
At low temperatures, equilibrium predicts that CH$_4$ and NH$_3$ would be present
and effective as greenhouse gases, and thus the published \citet{Lupu2014} model may overestimate the greenhouse effect and underestimate $T_{\rm eff}$ when surface temperatures are below 500 K (above which H$_2$O is abundant).
On the other hand, we are missing many opacity sources that would be present at higher temperatures, and thus in all likelihood the
model underestimates the greenhouse effect when surface temperatures are above 1000 K.
In practice we address the consequences of uncertain opacities by varying the total pressure of the atmosphere.
 
New calculations of $T_{\rm eff}$ for cloud-free runaway greenhouse H$_2$O-CO$_2$-N$_2$ atmospheres
\citep{Goldblatt2013,Kopperapu2013} find that 
$T_{\rm eff}=265$ during the runaway greenhouse phase of
planetary cooling
We expect that a deep CO$_2$ atmosphere will remain after the ocean of water condenses.
This state was not addressed explicitly by \citet{Lupu2014}.
We therefore retrofitted a smooth transition to the $\sim\!500$ K surface temperature expected under 100 bars of CO$_2$ \citep{Zahnle2006}.
What happens to the CO$_2$ at later times on longer timescales is discussed elsewhere \citep{Sleep2001,Sleep2014}.

\subsection{Physical parameters} 
\label{Physical parameters} 
 
For present purposes a ruthlessly simplified model of the Earth is the best choice.
We characterize the Earth by a surface temperature $T_{s}$ and an interior temperature $T_{i}$. 
In effect $T_{i}$ is the potential temperature, the temperature that every parcel in an adiabatic mantle would have
were it brought to the surface. 
We also assume that the mantle is uniform, in the sense that it has the same viscosity at all depths
and that it freezes everywhere at the same time. 
This assumption is roughly equivalent to assuming that the adiabat and the melting curve are parallel.
 We take $T_{\rm liq}=1800$ K and $T_{\rm sol}=1400$ K. 
 These are arbitrary values; our results are insensitive to them.
 We assume that the melt fraction $\phi$ is linear in $T_{i}$ between $T_{\rm liq}$ and $T_{\rm sol}$,
 \begin{eqnarray}
\label{phi}
\phi \!&\! =1\phantom{\left(T_{i}-T_{\rm sol} \right) /\left( T_{\rm liq}-T_{\rm sol} \right)} &\qquad T_{i}>T_{\rm liq} \nonumber\\
\phi \!&\!\!\!\! = \left(T_{i}-T_{\rm sol} \right) /\left( T_{\rm liq}-T_{\rm sol} \right) &\qquad T_{\rm liq}>T_{i}>T_{\rm sol} \nonumber\\
\phi \!&\! = 0\phantom{\left(T_{i}-T_{\rm sol} \right) /\left( T_{\rm liq}-T_{\rm sol} \right)}&\qquad T_{\rm sol} > T_{i}.
\end{eqnarray}
For $T_{i}>T_{\rm liq}$ and $T_{i}<T_{\rm sol}$, the heat capacity of the Earth is 
\begin{equation}
 C_{v\oplus}M_{\oplus} = C_{v,\rm man} M_{\rm man} + C_{v,\rm cor} M_{\rm cor} = 6.2\times 10^{30}{\rm~J/kg/K},
 \end{equation}
 in which the masses of the Earth, the mantle, and the core are $M_{\oplus}$, $M_{\rm man}$ and $M_{\rm cor}$,
 respectively.
The heat capacity of silicate is approximated by 
$C_{v,\rm man}=1200$ J/kg/K and that of iron by $C_{v,\rm cor}=650$ J/kg/K. 
The iron core does not freeze on the time scale of interest.
While silicates freeze, the effective heat capacity of the Earth is 
 \begin{equation}
 \label{CMprime}
 C^{\prime}_{v\oplus}M_{\oplus} = C_{v\oplus}M_{\oplus} + {Q_mM_{\rm man}\over T_{\rm liq}-T_{\rm sol}} = 1.0 \times 10^{31} {\rm~J/kg/K}
 \end{equation}
where the heat of fusion is $Q_m=4\times 10^5$ J/kg.  
With the above simplifications, the relation between ${\dot E_i}$ and $\dot{T_{i}}$ is 
\begin{eqnarray}
\label{edot}
{\dot E_i} \,=&\!\! C_{v\oplus}M_{\oplus} \dot{T_{i}} &\qquad T_{i}>T_{\rm liq},\,\, T_{\rm sol}>T_{i} \nonumber\\
{\dot E_i}  \,=&\!\! C^{\prime}_{v\oplus}M_{\oplus} \dot{T_{i}}  &\qquad T_{\rm liq}>T_{i}>T_{\rm sol} .
\end{eqnarray}
Including the atmosphere, the total ${\dot E}$ is
\begin{equation}
\label{total_edot}
{\dot E} = {\dot E_i} + C_{v,{\rm atm}}M_{\rm atm} \dot{T_{s}},
\end{equation}
in which $\dot{T_{s}}$ is related to $\dot{T_{i}}$ by the equations of parameterized convection described in the next section.
The thermal energy in the atmosphere is a small term.   

\subsection{Parameterized convection} 
\label{Parameterized convection} 

We follow previous workers by using parameterized convection to link
the cooling rate to the temperature and heat generation inside the Earth \citep{Solomatov2009,Lebrun2013}. 
In parameterized convection, the geothermal heat flow $F$
\begin{equation}
\label{Heat_flow}
F = C_0 {k_c \left(T_{i} -T_{s} \right) \over L}  Ra^n
\end{equation}
is related to the size and viscosity of the interior and the temperature gradient across
the surface boundary layer through the Rayleigh number $Ra$, 
\begin{equation}
\label{Rayleigh}
Ra = {\alpha_v g \left(T_{i} -T_{s} \right) L^3 \over \kappa \nu} .
\end{equation}
 Equations \ref{Heat_flow} and \ref{Rayleigh} can be combined to give a closed-form expression for $F(T_s,T_i)$,
\begin{equation}
\label{combined}
F = C_0 k_c \left(T_{i} -T_{s} \right)^{1+n} L^{3n-1} \left( {\alpha_v g \over \kappa\, \nu (T_i) }   \right)^{n} .
\end{equation}
We use $k_c=500$ J/m$^2$/s/K for the thermal conductivity;
$\alpha_v$ is the volume thermal expansivity,
$g$ the gravity, and $\kappa$ is the thermal diffusivity ($\kappa\equiv\! k_c/C_{v,{\rm man}}/\rho$).
We will use $\alpha_v = 2.4\times 10^{-5}$ K$^{-1}$, $\kappa = 10^{-6}$ m$^2$/s, and $g=10$ m/s$^2$.
The characteristic distance scale $L$ is usually identified with the thickness of the mantle. 
In the soft turbulent regime we take $C_0=0.089$ and $n=1/3$ \citep{Solomatov2009}. 
With $n=1/3$, $L$ cancels out, so that the heat flow is determined only by the properties of the boundary layer.

\begin{figure}[!htb]
   \centering
\includegraphics[width=1.0\textwidth]{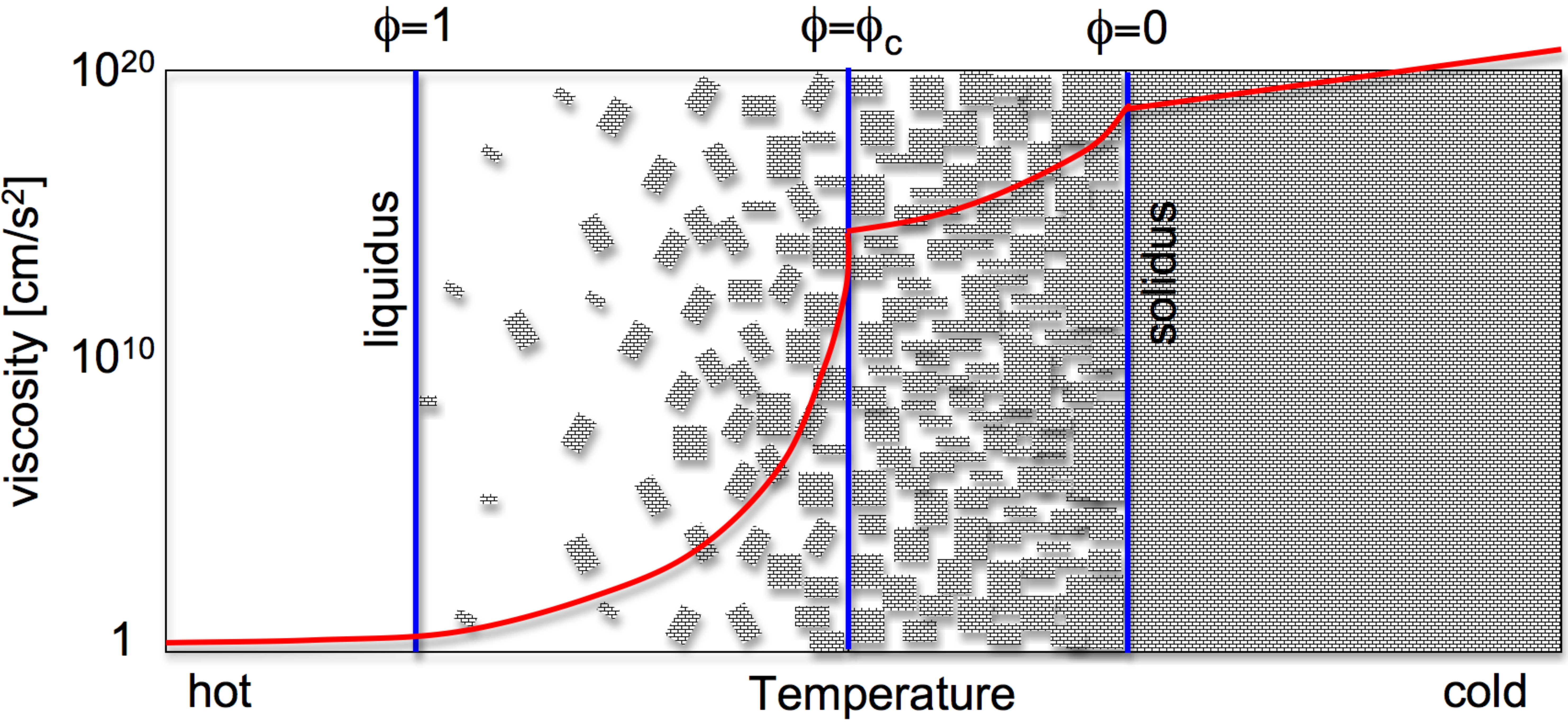} 
\caption{Semi-quantitative cartoon illustrating the rheological transition in a cooling magma ocean.
Cooling proceeds from left to right. At the liquidus refractory crystals begin to freeze out.  
With further cooling the crystal fraction increases and the liquid fraction $\phi$ decreases.
At the rheological transition $\phi=\phi_c\approx 0.4$. 
At the solidus the melt fraction $\phi=0$.
The corresponding kinematic viscosity $\nu$ is shown on a logarithmic scale.
The discontinuity in $\nu$ at $\phi=\phi_c$ ($T_i=T_{\rm crit}$) corresponds to a melting point.}
 \label{fig:Figure1}
\end{figure}

The kinematic viscosity $\nu$ (m$^2$/s) is a very strong function of temperature $T_i$ and therefore of central importance to convection and tidal heating.
We follow \citet{Abe1997,Solomatov2009,Lebrun2013} by treating the temperature dependence of viscosity in two regimes,
the hotter resembling a liquid and the cooler resembling a solid, the regimes divided by $T_{\rm crit}$.  
The phase transition is a sharp function of the fraction $\phi$ of silicate that is liquid.
In the liquid, \citet{Solomatov2009} parameterizes viscosity by 
\begin{equation}
\label{liquid_viscosity}
\nu\left(T_{i}\!>\!T_{\rm crit}\right) = 6\times 10^{-8} \exp{\left({4600{\rm ~K}\over T_{{\rm i}}-1000{\rm ~K}}\right)} \left({1-\phi_{\rm crit}\over \phi -\phi_{\rm crit}}\right)^{2.5} \quad{\rm m}^2/{\rm s}.
\end{equation}  
This expression implies that $\nu\rightarrow\infty$ as $\phi\rightarrow\phi_{\rm crit}$; i.e., it is singular at $\phi=\phi_{\rm crit}$.
In the solid, 
\begin{equation}
\label{solid_viscosity}
\nu\left(T_{i}\!<\!T_{\rm crit}\right) = 1\times 10^{15} \exp{ \left( {T_{\rm sol}-T_{{\rm i}}\over 58{\rm ~K}} \right) } \exp{\left(-\alpha_{\phi} \phi \right)}  \quad{\rm m}^2/{\rm s},
\end{equation}
with $\alpha_{\phi}=26$ \citep{Solomatov2009}.
\citet{Abe1997} places the boundary at $\phi = \phi_{\rm crit} = 0.4$.  
For $\phi_{\rm crit} = 0.4$, $T_{\rm crit}= 1560 {\rm ~K}$.
The particular values of $\phi_{\rm crit}$ and $T_{\rm crit}$ are unimportant.
Figure \ref{fig:Figure1} illustrates the situation.

\subsection{Moonless Earths}
\label{Moonless}

\begin{figure}[!htb]
   \centering
   \includegraphics[width=1.0\textwidth]{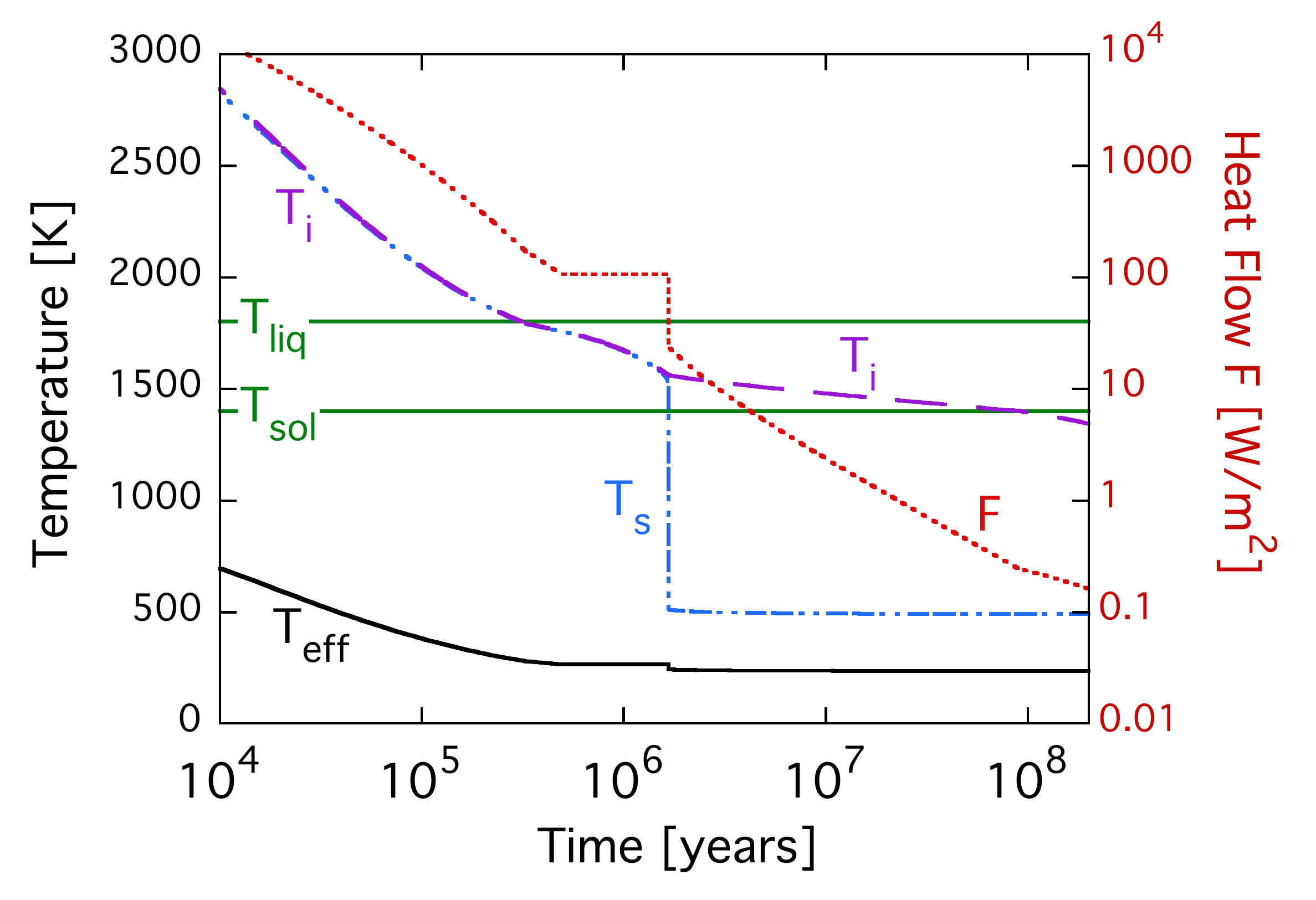} 
   \caption{Thermal evolution of an Earth after a non-moon-forming giant impact of the same energy as
   the canonical Moon-forming impact.
   Results are shown for a 100 bar BSE atmosphere.
   Internal temperature $T_{i}$ (mantle), surface temperature $T_{s}$, and effective radiating temperature $T_{\rm eff}$ are plotted
   against the left-hand axis.  The liquidus and solidus are indicated.
   Geothermal heat flow $F$ (dotted curve) is plotted against the right-hand axis.
   The heat flow plateaus at $F\sim\! 110$ W/m$^2$ while the Earth passes through the runaway greenhouse, 
   a phase that lasts from $\sim\! 0.5$ Myr to $\sim\! 1.7$ Myr.
   This period corresponds to a liquid mantle.
   It ends when the mantle becomes too viscous to sustain $F \sim\! 110$ W/m$^2$,
   at which point $T_s$ drops to low values, here $\sim$500 K under 
   $\sim 100$ bars of CO$_2$. The mantle remains partially molten for $\sim100$ Myrs.
   The latter is what is often meant by ``magma ocean.''}
   \label{fig:Figure2}
\end{figure}

Impacts on the scale of the Moon-forming impact need not create a moon.
It is illustrative to consider a case without the Moon for comparison with previous work 
\citep{Hamano2013,Lebrun2013}.
Computed temperatures and heat flow histories for a 100 bar BSE-equilibrated atmosphere are shown in Figure \ref{fig:Figure2}.
In these simulations the initial temperature is set at $T_i=3500$ K at $t=100$ years.

Cooling takes place in two sharply defined regimes (Figure \ref{fig:Figure2}).
In the first regime the three leading terms in the planet's energy budget,
\begin{equation}
\sigma T_{\rm eff}^4 = \left(1-A\right) F_{\odot}/4 + F,
\end{equation}
are of comparable magnitude, 
with the geothermal heat flow $F\sim 110$ W/m$^2$ making up the difference between 
net insolation $\sim\!170$ W/m$^2$ and radiative cooling at the runaway greenhouse threshold $\sim\!280$ W/m$^2$.
In this regime the temperature contrast $\Delta T=T_i-T_s$ across the surface thermal
boundary layer is small compared to $T_i$.
This lasts about 1 million years.
The transition to the second regime takes place when the mantle's
viscosity grows too great to sustain $F\sim 110$ W/m$^2$, 
at which point the mantle undergoes the phase transition depicted in Figure \ref{fig:Figure1}.
After the phase transition heat flow is negligible (here $\sim 20$ W/m$^2$ at first, but falling off fairly quickly)
and the surface temperature is determined by
insolation and the greenhouse effect of $\sim$100 bars of CO$_2$. 

As \citet{Matsui1986} and \citet{Hamano2013} point out, the value $F\approx 110$ W/m$^2$ is particular to  
ancient Earth with insolation $F_{\odot}=1.0\times 10^3$ W/m$^2$ and albedo $A=0.3$. 
At Venus's distance from the Sun, 
the heat flow required to support a runaway greenhouse atmosphere with $A=0.4\pm0.1$
is of the order of $F=10\pm 40$ W/m$^2$.
Venus after the impact could have been held in a state with a molten surface
for tens of millions of years or more, and if its albedo were less than $A\approx 0.38$ it 
need not have emerged from this state until after its water was lost to space \citep{Hamano2013}.  

\section{The Moon and Tides}

\label{Tides}

Tidal heating by the nearby Moon was a major term in early Earth's energy budget. 
The heat comes from braking the rotation of the Earth.
Tidal dissipation is usually described by the $Q$ parameter, 
 defined as the fraction of energy in the wave that is damped out during a single 
period \citep{GoldreichSoter1966}.   
The value of $Q$ can often be deduced empirically while presupposing nothing about the underlying physics
of tidal dissipation.
%
We will use $Q$, but for the congealing mantle we will assume that
tidal dissipation can be described by viscosity.
This is a reasonable approximation as the mantle starts to freeze because 
ordinary viscous dissipation would be big \citep{Moore2003}.
Where viscous dissipation dominates, $Q$ can be computed (i.e., instead of assuming $Q$ directly, we assume 
a particular physical mechanism for dissipation and a particular description of how viscosity evolves). 
But at very early times when the mantle is effectively inviscid, or at modern times when the mantle
is solid, ordinary viscous dissipation
is probably negligible and some other source of dissipation will likely be more important.
In a hot molten Earth we use observed values of $Q$ in fluid planets as a guide.
For modern times we assume an asymptotic viscosity that brings the Earth-Moon system to its current
state in 4.5 Ga,
it being understood that the bulk of tidal dissipation on the Earth today actually takes place at the interface between
the solid surface and the liquid water oceans \citep{Lambeck1980}.

\subsection{Simple tidal dissipation}

We assume that, locally, viscous tidal dissipation scales as the viscosity multiplied by the square of the strain rate ${\dot \epsilon}$.
The strain $\epsilon$ is the fractional (dimensionless) deformation induced by the tide.
It is of order of $H/R_{\oplus}$, where $H$ is the amplitude of the tide.
The rate of strain is therefore on the order of ${\dot \epsilon} = Hf/R_{\oplus}$,
where $f$ is the forcing frequency of the tide.  
Viscous heating per unit volume scales as $\rho\nu{\dot \epsilon}^2$, which for a uniform planet equates
to $M_{\oplus}\nu{\dot \epsilon}^2$ as a whole.
The peak-to-peak amplitude of the lunar semidiurnal equilibrium tide is
\begin{equation}
\label{equil_tide}
H = {3 \over 2} h_2 R_{\oplus}\left({M_{\footnotesize\leftmoon}\over M_{\oplus}}\right)\left({R_{\oplus}\over a}\right)^{3},
\end{equation}
in which $M_{\footnotesize\leftmoon}$ is the mass of the Moon, $a$ is the semimajor axis of the Moon's orbit, and
 $h_2$ is the amplitude Love number \citep{Kaula1968}, 
\begin{equation}
\label{h2}
h_2 = {5\over 2} \left( 1+ {19 \over 2} {\mu \over \rho g R_{\oplus}}\right)^{\!-1} .
\end{equation}
The dimensionless quantity $(19/2)(\mu/\rho g R_{\oplus})$ 
compares the material rigidity $\mu$ (resistance to shear) to the weight of the Earth.
The angular frequency of the semidiurnal lunar
tide --- the only tide we will consider here --- is $f=2\Omega -2n_1$,
where $n_1$ denotes the angular velocity of the Moon in its orbit.

Viscous heating must be modified to take into account how quickly matter can respond to forcing.
The resulting heuristic expression\footnote{According to \citet{Makarov2014}, this is exact if multiplied by a factor of $19/25$.} for tidal dissipation in Earth is 
\begin{equation}
\label{Edot_tide}
{\dot E}_{\rm tide} \approx {M_{\oplus}\nu {\dot \epsilon}^2 \over 1 + f^2\tau_{\nu}^2},
\end{equation}
where $\tau_{\nu}$ represents the viscous relaxation time scale,
which  is best written (M.\ Efroimsky, personal communication) in the form 
\begin{equation}
\label{inverse_tau_squared}
\tau_{\nu} = \tau_{\rm M} + \tau_{\rm D} 
\end{equation}
with $\tau_{\rm M} \equiv \rho\nu/\mu$ being the Maxwell time scale
and $\tau_{\rm D} \equiv 19 \nu/2gR_{\oplus}$ being 
Darwin's time scale for relaxation of a viscous globe under its own gravity \citep[][p.\ 641]{Lamb1932}.
Expressions in the form of Eq.\ \ref{Edot_tide} are often encountered in the planetary literature to describe tidal dissipation
in a variety of objects \citep[e.g.,][]{Moore2003,Garrick-Bethell2010}.
For the shear modulus $\mu$ we assume 
\begin{eqnarray}
\label{rigidity}
\mu =&\!\!\! 0\phantom{.3\times 10^{11}\exp{\left(-\alpha_{\phi} \phi \right)}} &\quad T_{i}>T_{\rm crit} \nonumber\\
\mu =&\!\!\! 1.3\times 10^{11} \exp{\left(-\alpha_{\phi} \phi \right)}  &\quad T_{\rm crit}>T_{i}>T_{\rm sol} \nonumber\\
\mu =&\!\!\!1.3\times 10^{11}  \phantom{\exp{\left(-\alpha_{\phi} \phi \right)}} &\quad T_{\rm sol} > T_{i}
\end{eqnarray}
where the units of $\mu$ are kg/m/s.  
The shear modulus is related to the Love numbers $h_2$ (Eq \ref{h2}) and $k_2$ (Eq \ref{k2}).
For Earth today, \citet{Kaula1968} and \citet{Lambeck1980} give $h_2\approx 0.6$ and $k_2 \approx 0.3$.
These values imply that $\mu = 1.3\times 10^{11}$ kg/m/s can be used as an effective average rigidity of the Earth today.
The presumed temperature dependence of $\mu$ in Eq.\ \ref{rigidity} is arbitrary. 

Viscous dissipation can be expressed in terms of $Q$.
\citet[][Eq.\ 2.3.1]{Kaula1968} defines $Q$ as
\begin{equation}
 {2\pi \over Q} \equiv {1\over E} \oint {\dot E} dt 
 \end{equation}
where $E$ represents the energy in the wave and the integral is over one full cycle. 
This is equivalent to writing
\begin{equation}
\label{Kaula}
Q = {\dot E}_{\rm pot}/{\dot E}_{\rm tide},
\end{equation}
where the total potential energy in the tidal wave is \citep[][p.\ 201]{Kaula1968}.
\begin{equation}
\label{Epot}
{\dot E}_{\rm pot} = {3\over 4} k_2 G M_{\footnotesize\leftmoon}^2 {R_{\oplus}^5 \over a^6} f,
\end{equation}
where the potential Love number is 
\begin{equation}
\label{k2}
k_2 = {3\over 2} \left( 1+ {19 \over 2} {\mu \over \rho g R_{\oplus}}\right)^{-1}.
\end{equation}
We use Eq.\ \ref{Kaula} to generate $Q_{\oplus}$.

At temperatures below the solidus, for which viscosity is very large, ordinary viscous dissipation 
at tidal frequencies is probably unimportant.
 To bring the Moon to its present orbit at the present time requires $Q_{\oplus}\sim 30$, which
 we obtain by setting an upper bound on
 ``viscosity'' in Eq.\ \ref{Edot_tide} of $\nu_{\rm max} = 1.3\times 10^{13}$ m$^2$/s.
 We apply the upper bound on $\nu$ only to tidal dissipation in Eq \ref{Edot_tide}.
 We do not apply it to the Rayleigh number and related quantities pertinent to parameterized convection.
 
At high temperatures the ordinary viscosity of mantle liquids is expected to be very small. 
 With $\nu=10^{-4}$ m$^2$/s, we estimate that
  $Q_{\oplus}\sim\sigma_2\tau_2 \sim 10^{14}$,
 where $\sigma_2=4g/5R_{\oplus}$ is the relevant natural oscillation frequency of a self-gravitating fluid globe \citep[][art.\ 262, Eq.\ 10]{Lamb1932}
 and $\tau_2=R^2_{\oplus}/5\nu$ is the relevant timescale for viscous dissipation in a self-gravitating fluid globe when viscosity is small \citep[][art.\ 355, Eq.\ 12]{Lamb1932}.
Convecting fluid planets may provide useful guidance.
 General constraints set by tidal evolution of the Galilean satellites  
 suggest that $Q$ of Jupiter is 
 between $10^5$ and $10^6$ \citep{GoldreichSoter1966}.
 Astrometric observations suggest that $Q$ of Jupiter is $\sim\! 3.5\times 10^4$ \citep{Lainey2009}
 and $Q$ of Saturn just $1400$ \citep{Lainey2012}, the latter remarkably low compared to the $7\times 10^4$ 
estimated by \citet{GoldreichSoter1966}.
 The uncertainty and variance reminds us that origin of dissipation in the giant planets is unknown.
 Similar arguments have been made for binary stars;
 these suggest that $Q$ is of the order of $10^5-10^6$ \citep{Meibom2005}.
 On the other hand, 
 at least one hot superJupiter---the 10 Jupiter mass WASP-18b, currently parked in a 0.94 day orbit---may require $Q>10^9$ \citep{Hellier2009}.

\subsection{Simple tidal evolution}

The simplest assumption is that the Moon evolved away from the Earth while maintaining a circular orbit.
We will make this assumption here.
The Moon revolves about the Earth at the angular frequency   
\begin{equation}
\label{omega}
n_1 = \sqrt{G\left(M_{\oplus}+M_{\footnotesize\leftmoon}\right)/ a^3}.
\end{equation}
The total Earth-Moon system angular momentum ignoring lunar spin and assuming a circular orbit is
\begin{equation}
L = I_{\oplus}\Omega + M_{\leftmoon} \sqrt{ G \left( M_{\oplus}+ M_{\footnotesize\leftmoon}\right) a } ,
\end{equation}
in which $\Omega$ denotes Earth's angular rotation rate.
The canonical model of the Moon-forming impact assumes that the angular momentum of the Earth-Moon system 
has been conserved.
With ${\dot L}=0$, the Moon recedes at the rate
\begin{equation}
\label{adot}
{\dot a} = {-2I_{\oplus}{\dot \Omega} a \over L - I_{\oplus}\Omega} .
\end{equation}
%
With ${\dot a}$ given by Eq \ref{adot},
the tidal dissipation is related to the rate of braking of Earth's spin by
\begin{equation}
{\dot E}_{\rm tide} = -\left( I_{\oplus}\Omega - { GM_{\oplus}M_{\footnotesize\leftmoon}\over  a} {I_{\oplus}\over L - I\Omega}\right) {\dot \Omega} ,
\end{equation}
in which $I_{\oplus}=0.33\,M_{\oplus}R_{\oplus}^2$ represents Earth's moment of inertia.
Note that ${\dot \Omega}$ is negative and ${\dot a}$ and ${\dot E}_{\rm tide}$ are positive.

\begin{figure}[htb]
   \centering
   \includegraphics[width=1.0\textwidth]{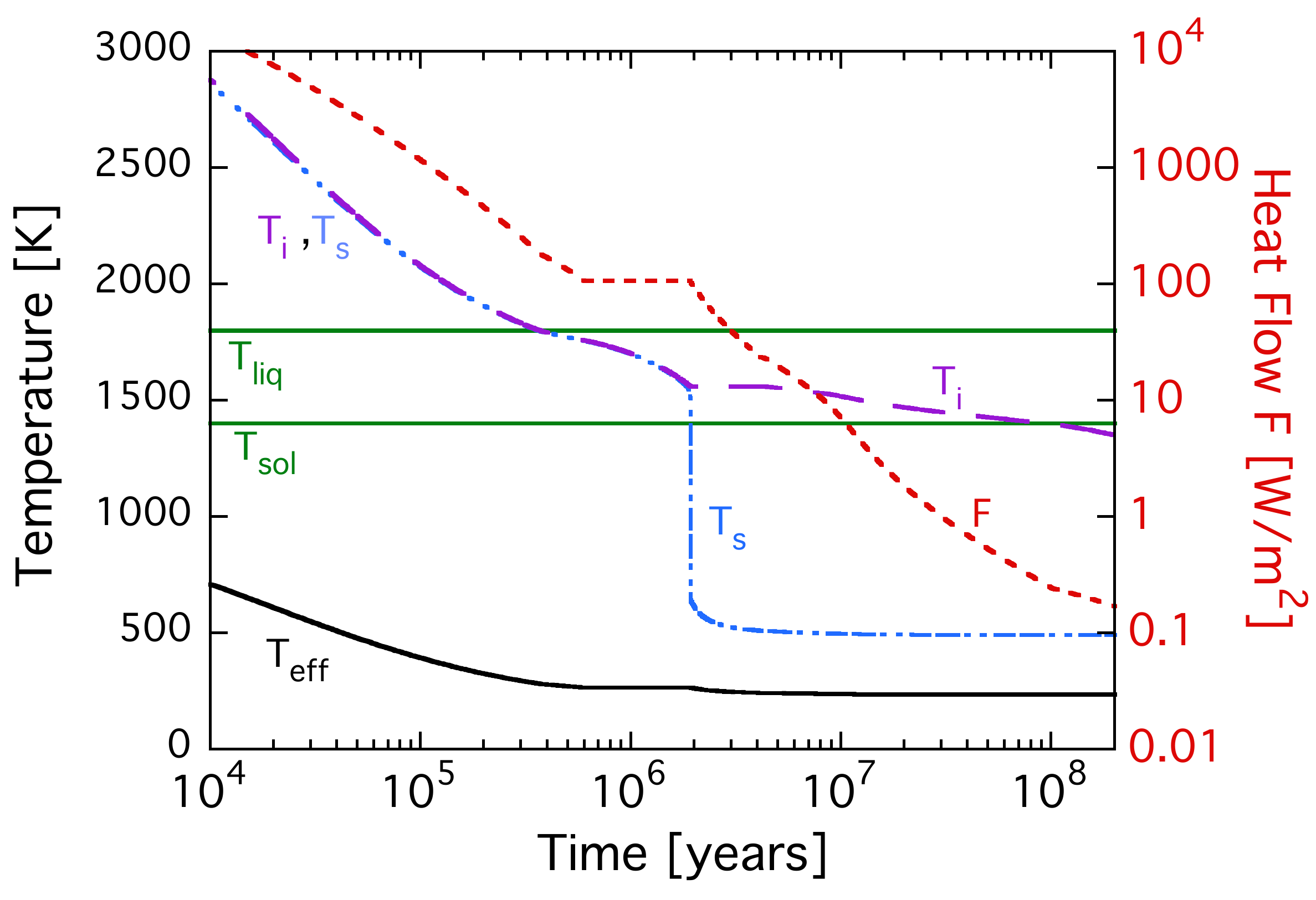} 
   \caption{Thermal evolution of the Earth after a canonical Moon-forming impact.
   Results are shown for the 100 bar BSE-equilibrated atmosphere.
   The results are very similar to Figure \ref{fig:Figure2}.
   Tidal heating extends the runaway greenhouse phase by a few hundred thousand years compared to the moonless case.
   Geothermal heat flow $F$ is about an order of magnitude higher than in the moonless case for a few tens of Myrs.}
   \label{fig:Figure3}
\end{figure}

Figure \ref{fig:Figure3} shows thermal evolution of an Earth after a canonical Moon-forming impact that 
forms a Moon.
Other particulars are the same as with the moonless impact of Figure \ref{fig:Figure2}.
Tidal heating extends the duration of the runaway greenhouse phase by a few hundred thousand years,
and greatly increases geothermal heat flow for tens of millions of years afterward,
but at first glance the differences between Figures \ref{fig:Figure2} and \ref{fig:Figure3} are subtle.

\begin{figure}[htb]
   \centering
   \includegraphics[width=1.0\textwidth]{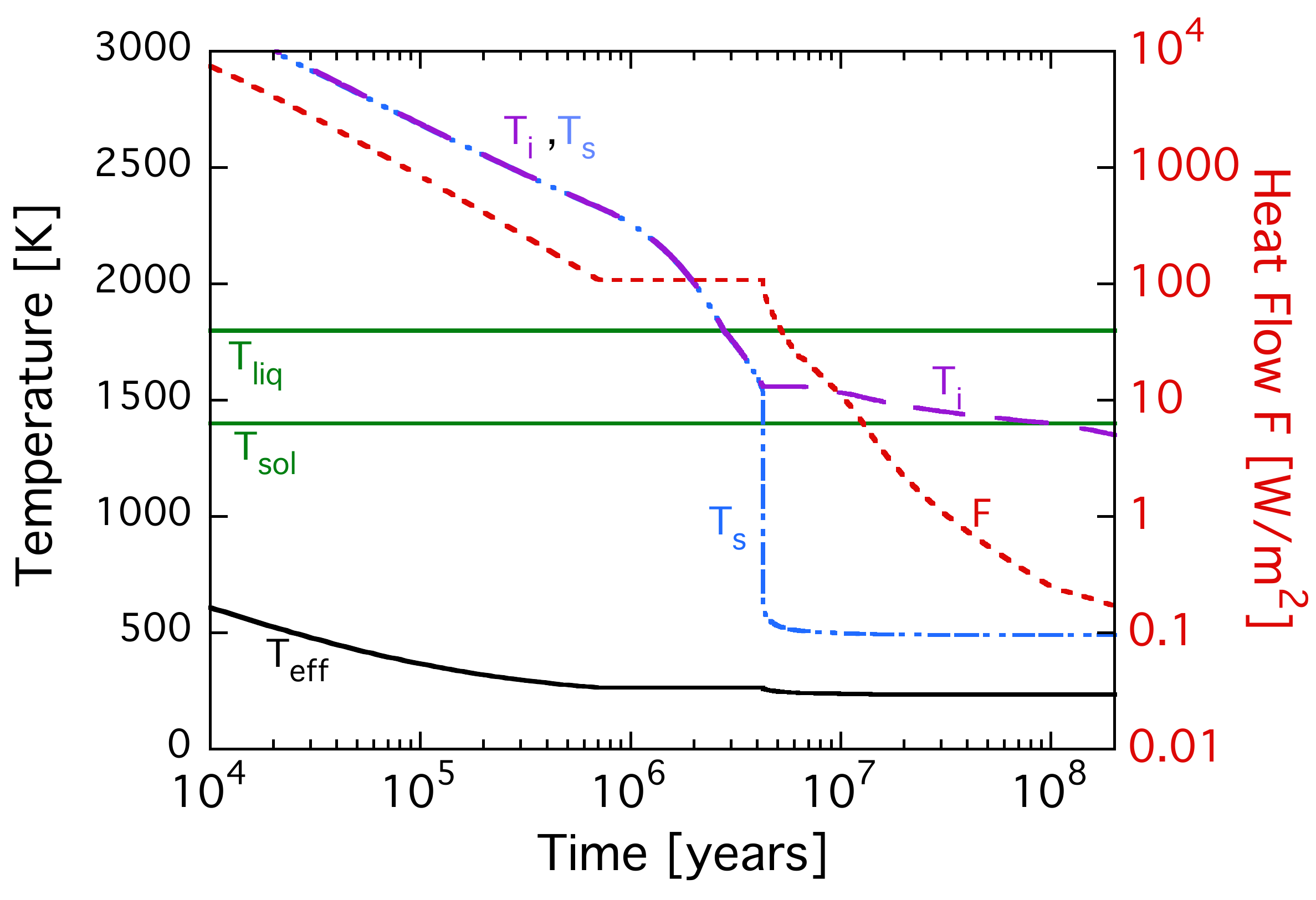} 
   \caption{Thermal evolution of the Earth after a canonical Moon-forming impact, with more volatiles.
   Results are shown for a 1000 bar BSE-equilibrated atmosphere.  
   The thicker atmosphere more than the doubles the duration of the runaway greenhouse phase,
   and increases the Earth's fully molten stage from 0.3 Myrs (Fig.\ \ref{fig:Figure3}) to 3 Myrs.    
    }
   \label{fig:Figure4}
\end{figure}

Changing the radiative properties of the atmosphere has greater effect than adding the Moon.
Figure \ref{fig:Figure4} shows a comparable thermal evolution under a 1000 bar post-impact atmosphere. 
This is toward the thicker end of the range we might expect for Earth,
but the thicker atmosphere also serves as a proxy for additional opacity sources that are undoubtedly present at very high temperatures
but are not included in our model, or a proxy for the greater opacity of an atmosphere that is more reduced than the BSE. 
The thicker atmosphere increases how long Earth remains liquid by an order of magnitude,
from $\sim 0.3$ Myrs for the 100 bar atmosphere to $\sim\! 3$ Myrs. 
The thicker atmosphere also doubles how long the Earth remains in the runaway greenhouse state (from $\sim\! 2$ to $\sim\! 4$ Myr).
Since the two cases start with the same amount of thermal and rotational energy,
 the differences between Figures \ref{fig:Figure3} and \ref{fig:Figure4} stem from how quickly the atmosphere radiates when it is much hotter than the runaway
greenhouse state.  This in turn is determined by the opacity of the atmosphere
when the surface is very hot, $T_s>2500$ K, a kind of atmosphere that is only now beginning to be studied.

\begin{figure}[htb]
   \centering
   \includegraphics[width=1.0\textwidth]{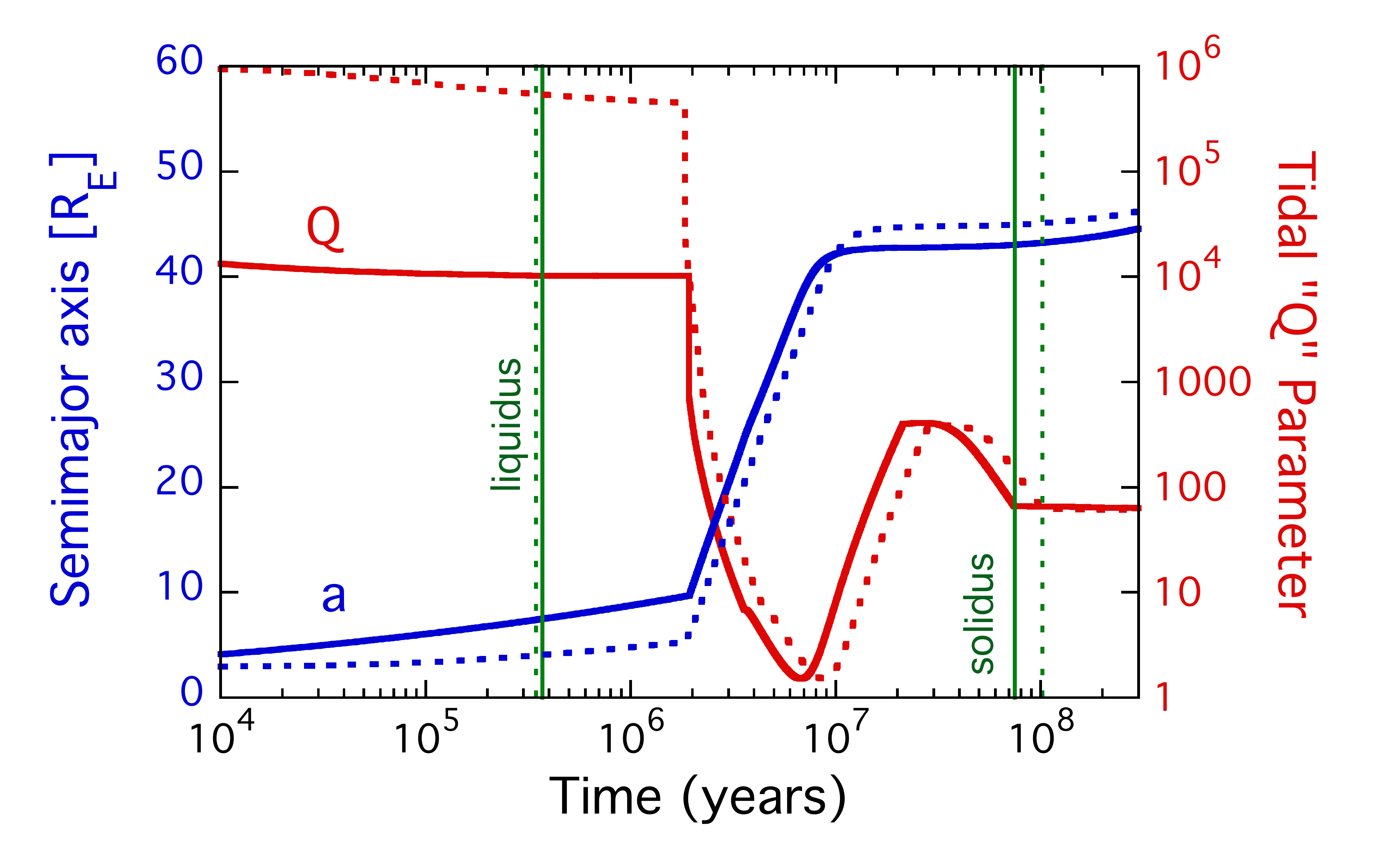} 
   \caption{The distance to the Moon $a$ and the tidal dissipation parameter $Q_{\oplus}$
   of the Earth after a canonical Moon-forming impact.
   Results are shown for high (dotted) and low (solid) values of $Q$ in the fully molten Earth.
   As it congeals, the mantle becomes extremely dissipative with $Q_{\oplus}$ near unity and the Moon's orbit evolves quickly.  
  After the mantle freezes the viscosity increases, $Q_{\oplus}$ increases, and the Moon settles down.
   Both models assume a relatively thin 100 bar BSE-equilibrated atmosphere.
   The 1000 bar atmosphere gives qualitatively similar results. }    
   \label{fig:Figure5}
\end{figure}

It is interesting to consider what these models predict for 
the evolution of the distance to the Moon and the parameter $Q_{\oplus}$ of the Earth.
Figure \ref{fig:Figure5}
shows two cases that differ only in the assumed upper bound on $10^4<Q_{\oplus}<10^6$ for the fully liquid Earth.
Both cases assume the Earth is enveloped in the 100 bar atmosphere.
  There are no striking differences between the two models.
   A minor difference is that the higher $Q_{\oplus}$ model delays tidal heating somewhat, thus extending the 
   lifetime of the magma ocean 
   (defined as $T_i>T_{\rm sol}$) from 70 to 100 Myrs.
   The 1000 bar atmosphere gives qualitatively similar results, 
   with a somewhat slower tidal evolution and a somewhat longer-lived magma ocean.

\subsection{The Evection and Eviction Resonances}

The evection resonance describes the state in which the perigee of the Moon's orbit precesses in one year \citep{Touma1998}.  
While the Moon is in the evection resonance,
its line of apsides remains nearly at a constant angle to the Sun,
 and thus the Sun's gravitational torque on the Earth-Moon system does not average out, but rather becomes cumulative. 
This enables transfer of angular momentum from the Earth-Moon system to the Earth's orbit about the Sun \citep{Cuk2012}.

The possible importance of the evection resonance to the history to the lunar orbit was pointed out
by \citet{Touma1998}.
They showed that, if the Earth-Moon angular momentum were conserved, the evection resonance would have been encountered
at $a\approx 4.6 R_{\oplus}$, and  
that capture by the resonance from a nearly circular orbit is likely if $da/dt< 1$ km/yr, but
unlikely if $da/dt> 10$ km/yr.
They compared these results to a conventional constant $Q$ model that predicts $da/dt \approx 150$ km/yr at $a\approx 4.6 R_{\oplus}$. 
\citet{Touma1998} found that if captured the Moon can evolve into a highly eccentric orbit with $e\approx 0.5$, 
which in their model is what allows the Moon to escape the resonance.

  \begin{figure}[!htb] 
   \centering
\includegraphics[width=1.0\textwidth]{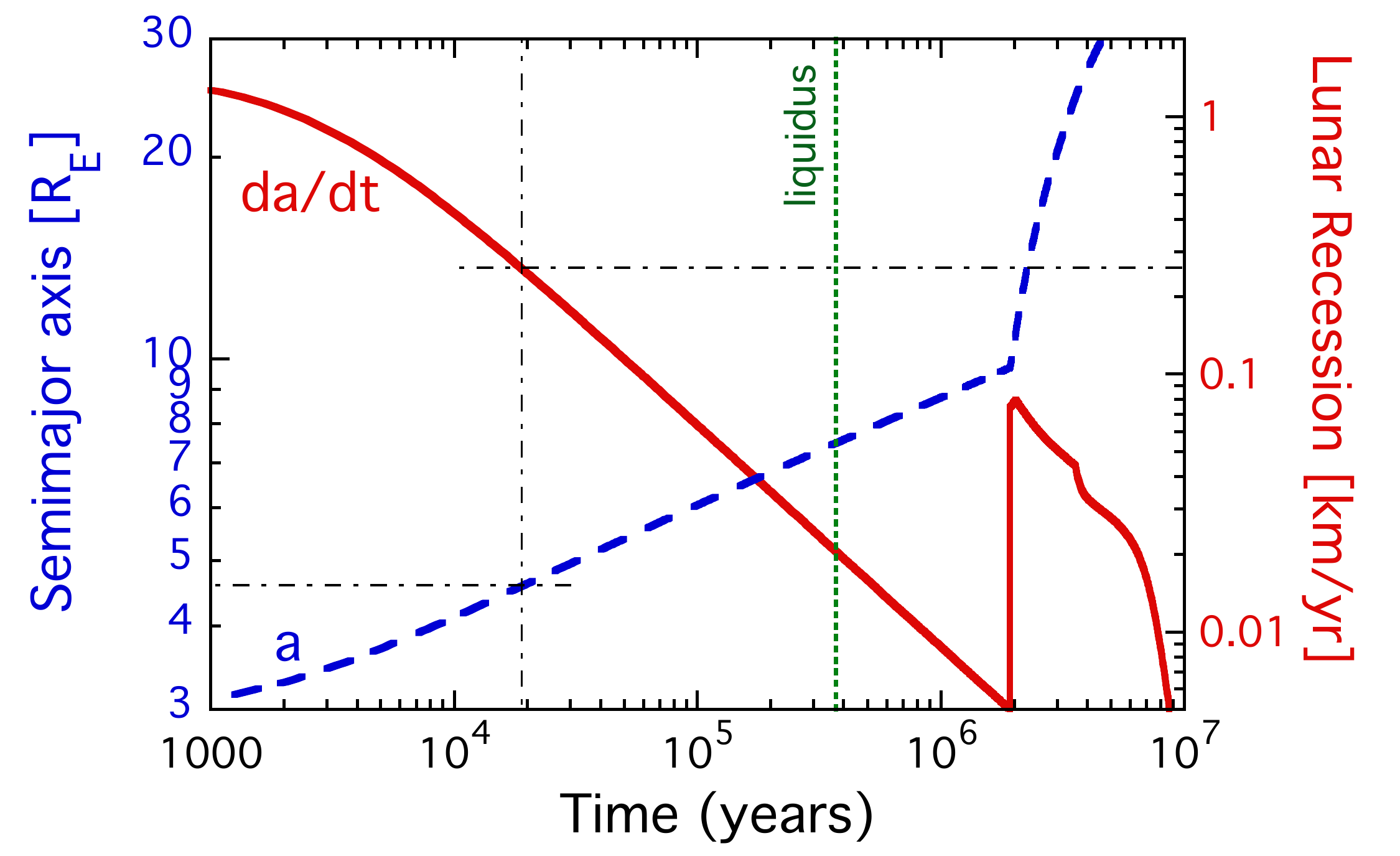} 
   \caption{Distance to the Moon $a$ and lunar recession rate $da/dt$ are plotted for our standard case
   with a 100 bar BSE atmosphere.
   \citet{Touma1998} show that the evection resonance is encountered at $a\sim 4.6 R_{\oplus}$,
   and that resonance capture from a near circular orbit requires $da/dt \leq 1$ km/yr.
   The dot-dashed lines indicate that the resonance at $4.6 R_{\oplus}$ is met early, after just $20,000$ Myrs
   (before the mantle has cooled to the liquidus),
   and that $da/dt< 0.3$ km/yr, the latter making resonance capture very probable.
    }
   \label{fig:Figure6}
\end{figure}
 
Integrations of the lunar orbit backward in time indicate that today's $5^{\circ}$ inclination to the ecliptic 
maps to a $\sim\!\!10^{\circ}$ inclination with respect to Earth's equator when the Moon was near the Earth \citep{Goldreich1966,Mignard1980}.
In the giant impact hypothesis, the Moon should have formed in Earth's equator and hence should have no inclination today.
\citet{Ward2000} proposed that a resonant interaction between the nascent Moon 
and the debris disk interior to it could generate an inclination of the right magnitude.
\citet{Touma1998} suggested that tidal evolution could raise the Moon's inclination if the Moon were captured successively,
first by the evection resonance, and then by a closely related ``eviction'' resonance.
The evection resonance gives the lunar orbit substantial eccentricity.
The eviction resonance can then give it substantial inclination. 
The chief challenge is that, to be captured by the eviction resonance,
the Moon must be evolving inward at the time (${\dot a}<0$), possible
only if the Moon were in a very eccentric orbit.
They find by numerical experiment that capture by the eviction resonance requires $\vert{da/dt}\vert< 0.2$ km/yr.
 
Figure \ref{fig:Figure6} illustrates that thermal blanketing by
Earth's atmosphere slows early tidal evolution of the Earth-Moon system by orders of magnitude compared to $da/dt = 150$ km/yr.  
The slow rate of tidal evolution makes resonance capture seem likely.
But we do not see much hope for the eviction mechanism here.
The problem is that the resonance encounters occur 
very early, when the Earth is still effectively inviscid, but not so early that the airless Moon is likely to be fully molten.
With tidal dissipation efficient in the Moon but inefficient in the Earth, the Moon's orbit
cannot acquire the substantial eccentricity that it would need to be captured by eviction.
We expect rather that the Moon does not acquire a substantial eccentricity until the Earth begins to freeze
and $Q_{\oplus}$ becomes small, which becomes the case
when $20R_{\oplus}<a<40R_{\oplus}$. 

\subsection{New Conventions}

Recently, \citet{Cuk2012} showed that, if the Earth-Moon system began with more angular momentum than it has today,
the evection resonance occurs further from the Earth (because the faster spinning Earth is more oblate), which makes
resonance capture more probable.
In their model the resonance remains occupied during a second contraction phase where both $e$ and $a$ slowly decrease.
They also showed that,
if the Earth-Moon system began with more angular momentum than it has today, it would lose that
angular momentum while captured in the evection resonance.

\begin{figure}[!htb]
   \centering
   \includegraphics[width=1.0\textwidth]{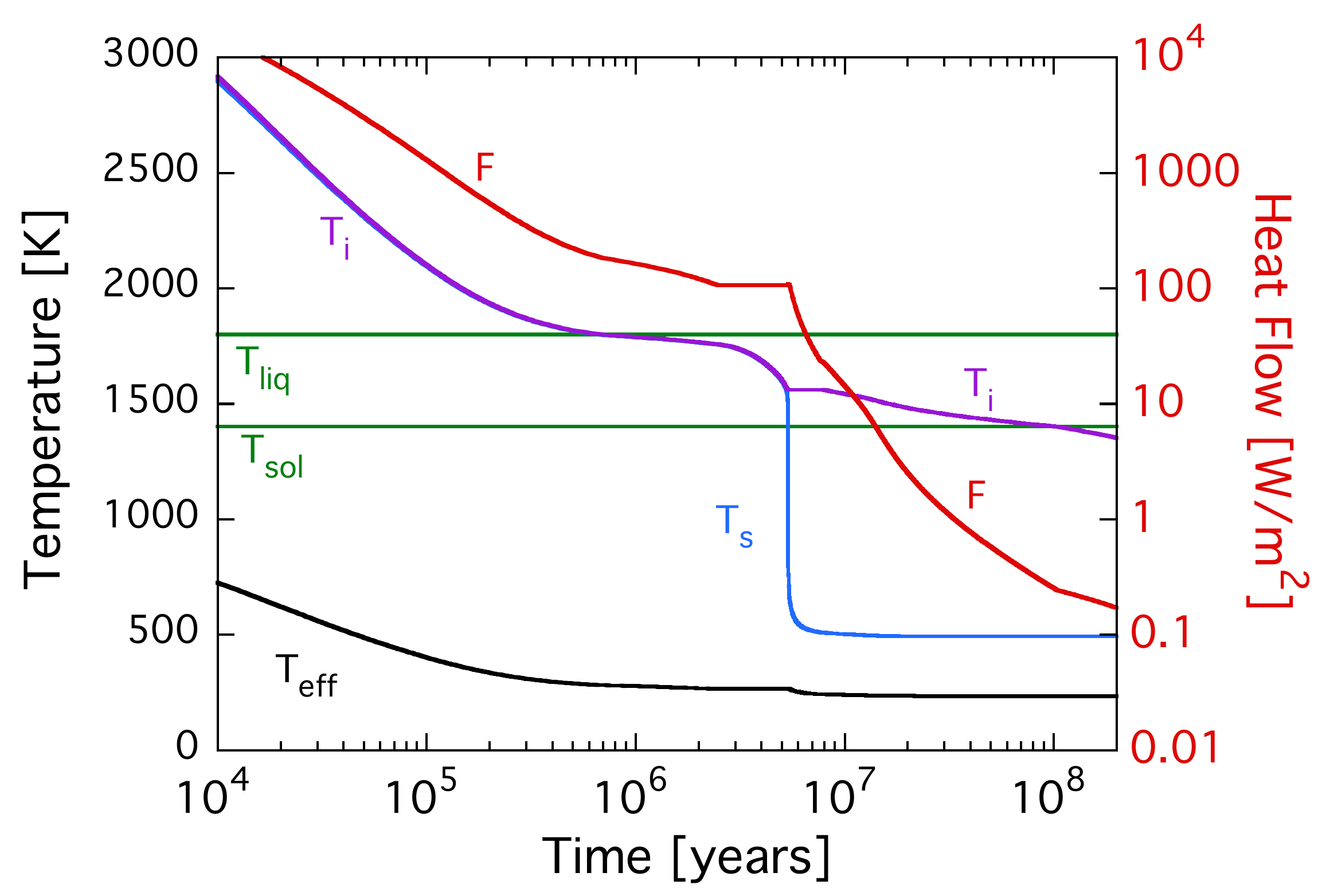} 
   \caption{Thermal evolution of the Earth after a high angular momentum Moon-forming impact.
   Results are shown for 100 bar atmospheres.
   Excess angular momentum in the Earth-Moon system is removed by the evection resonance, here crudely modeled by
   holding the Moon at the resonance distance until the angular momentum has been reduced to modern levels.
   This captures the spirit if not the details of {\'C}uk and Stewart's model.
   Compare to Figure \ref{fig:Figure3}.
  }
   \label{fig:Figure7}
\end{figure}

\begin{figure}[!htb]
   \centering
   \includegraphics[width=1.0\textwidth]{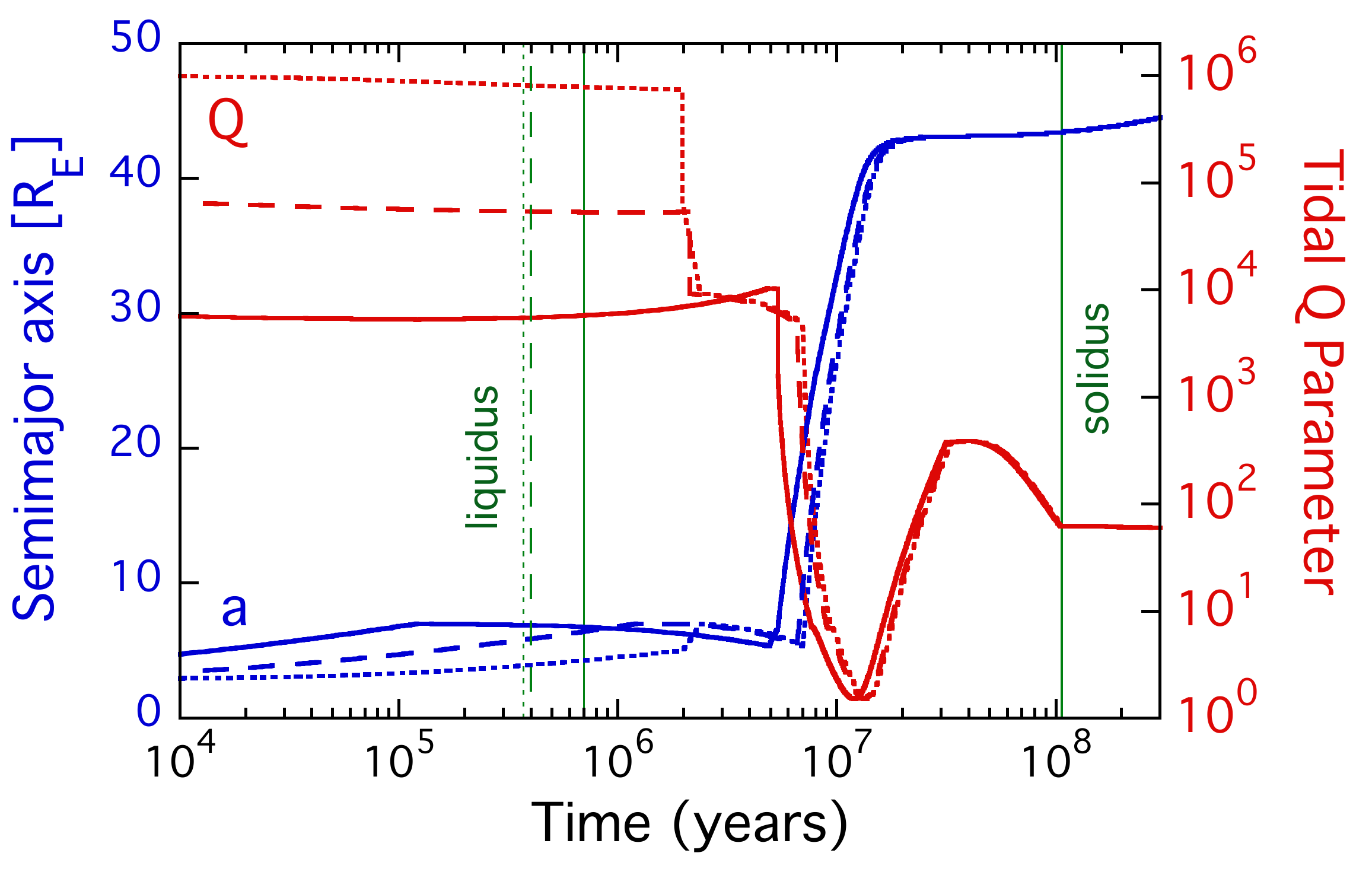} 
   \caption{Evolution of lunar distance $a$ and Earth's $Q_{\oplus}$ after high angular momentum Moon-forming impacts.
   Three cases are shown that correspond 
   to different upper bounds on $Q_{\oplus}$ in the fully molten Earth.
   All cases assume the 100 bar BSE-equilibrated atmosphere.
   Results can be compared to Figure \ref{fig:Figure5} for constant angular momentum.}
   \label{fig:Figure8}
\end{figure}

To approximate the evection mechanism at a level appropriate for this study,
we assume that when the Moon encounters the resonance at distance $a_r$,
it remains trapped while $L>L_{\rm now}$.
 Following \citet{Touma1998}, 
 we approximate the location of the evection resonance by
\begin{equation}
\left( {a_r \over R_{\oplus}} \right)^{7/2} = {3\over 2}J_{2} {n_1 \over n_2},
\end{equation}
in which $n_1$ (Eq.\ \ref{omega}) and $n_2$ are the angular velocities of the month and year,
and
\begin{equation}
J_2 = J_{20} \left( \Omega\,\, \over \Omega_{\circ} \right)^2
\end{equation}
approximates the oblateness of the Earth with $J_{20}=0.00108$.
 With these assumptions,
 \begin{equation}
 {\dot E_{\rm tide}} = -I_{\oplus} \Omega {\dot \Omega} - {1\over 2}M_{\footnotesize\leftmoon}n_1^2a_r{\dot a_r}
 \end{equation}
 and the angular momentum of the Earth-Moon system decreases as 
 \begin{equation}
 {\dot L} = I_{\oplus} {\dot \Omega} + M_{\footnotesize\leftmoon}n_1 a_r{\dot a_r}.
 \end{equation}
Figure \ref{fig:Figure7} shows thermal evolution of the Earth after a Moon-forming impact.
For Earth, the liquid magma ocean phase is prolonged to $\sim\! 6$ Myr by
the added heat released by tidal dissipation but otherwise resembles the angular momentum conserving case seen
in Figure \ref{fig:Figure3} above.

\begin{figure}[htb]
 \centering
   \includegraphics[width=0.95\textwidth]{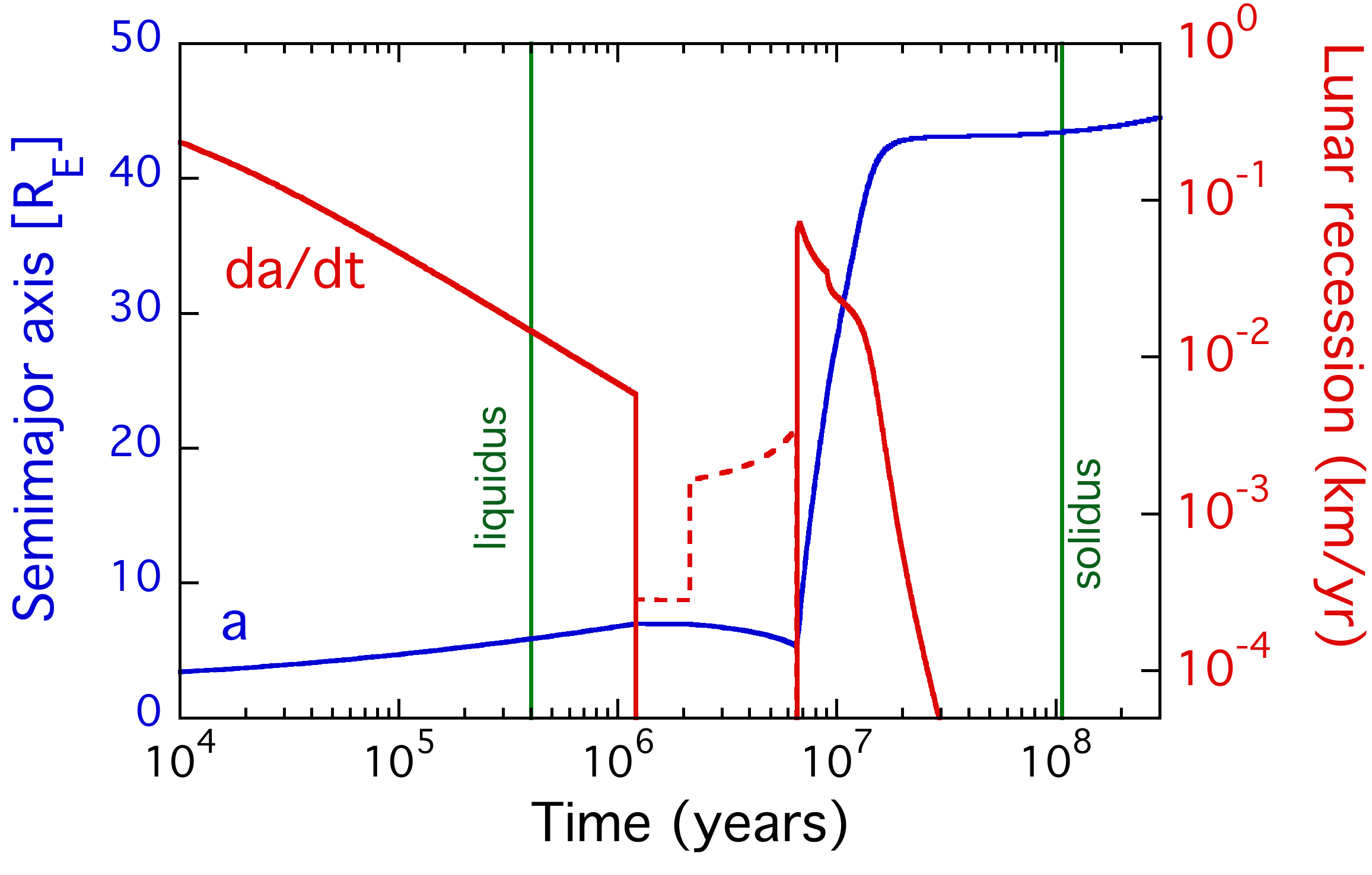} 
   \caption{Lunar recession velocities for a high angular momentum Moon-forming impact. 
   The case shown is the middle one from Figure \ref{fig:Figure8}.
   In resonance the Moon approaches the Earth ($da/dt<0$ indicated by dotted curve). 
   Earth encounters the evection resonance before it begins to soldify.
   Recession rates are low enough that resonance capture is certain.
   Details in the structure of $da/dt$ reflect changes in the mantle's rheological properties as it cools.
}
   \label{fig:Figure9}
\end{figure}

The evolution of $Q_{\oplus}$ in the Earth is the topic of Figure \ref{fig:Figure8},
the analog to Figure \ref{fig:Figure5}
for a high angular momentum Moon-forming impact. 
Figure \ref{fig:Figure9} shows lunar recession velocities for the middle case from 
 Figure \ref{fig:Figure8}.

\begin{figure}[htb]
 \centering
   \includegraphics[width=0.88\textwidth]{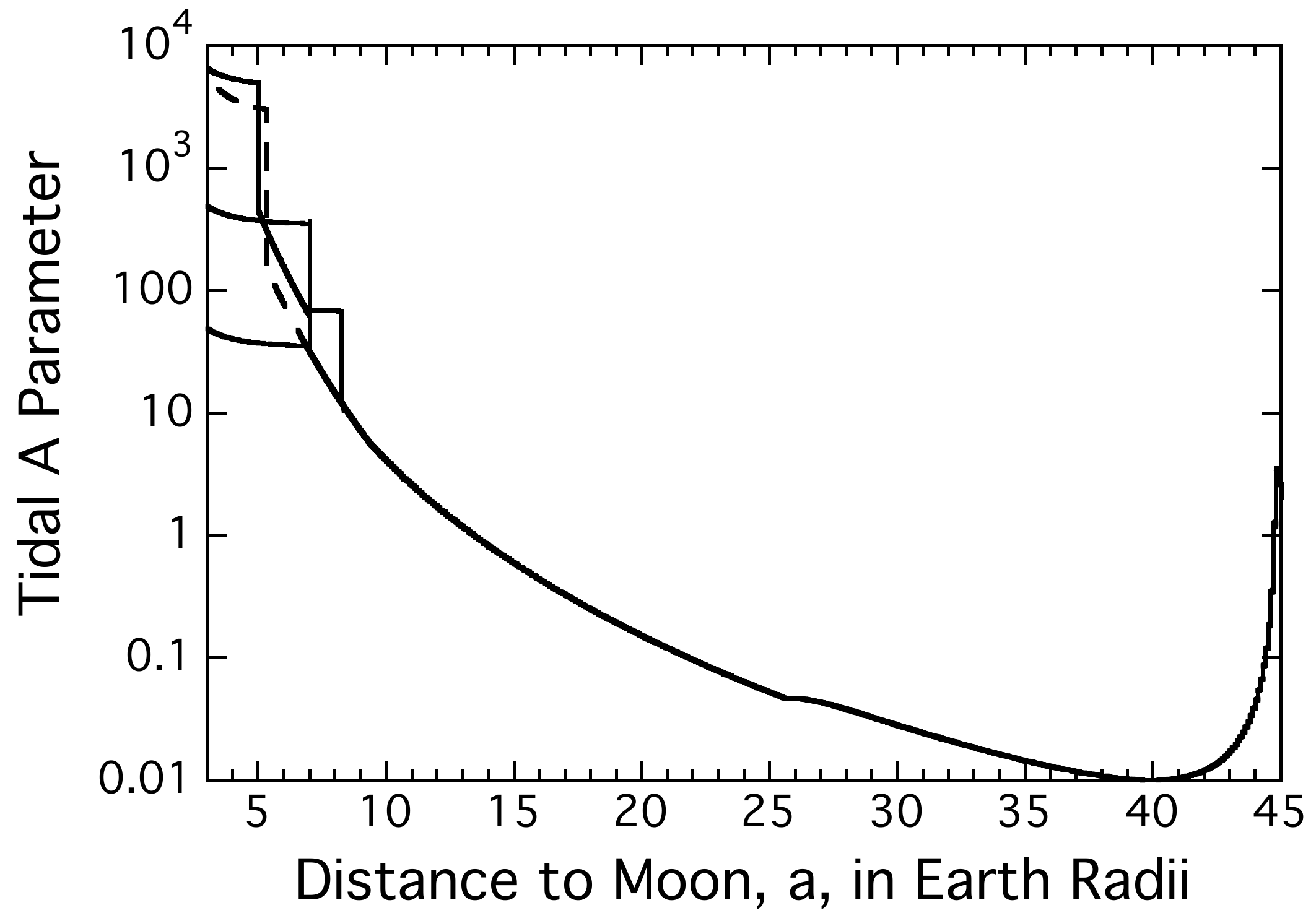} 
   \caption{Evolution of the $A$ factor that parameterizes the relative efficiency of tidal dissipation in the Moon
   to tidal dissipation in the Earth (Eq \ref{A-factor}).
   Lunar parameters $k_{2\footnotesize\leftmoon}=0.02$ and $Q_{\footnotesize\leftmoon}=20$ are held fixed.
   Several cases are shown: the conventional case from Fig.\ \ref{fig:Figure5} (dashed line),
   the three ``evection'' cases from \ref{fig:Figure8}, and another with a 1000 bar atmosphere (solid lines).
   Differences at the beginning stem from different assumptions about $Q_{\oplus}$ in the fully molten Earth.
   Otherwise the course of $A(a)$ is the same in all our models.
   In all cases the best prospects for growing a lunar eccentricity is with $A\ll 1$, when the distance to the Moon was
   between 20 and 40 $R_{\oplus}$.   
}
   \label{fig:Figure10}
\end{figure}

In general, tides in the Earth increase the eccentricity of the lunar orbit
while tides in the Moon decrease the eccentricity of the lunar orbit \citep{GoldreichSoter1966}.
Recent discussions of lunar orbital evolution have parameterized this struggle as an $A$ factor
 \citep{Mignard1980,Touma1998,Cuk2012},
\begin{equation}
\label{A-factor}
A \equiv {M_{\oplus}^2\over M_{\footnotesize\leftmoon}^2}{R_{\footnotesize\leftmoon}^5 \over R_{\oplus}^5} { k_{2\footnotesize\leftmoon}\over Q_{\footnotesize\leftmoon}} {Q_{\oplus}\over k_{2\oplus}} 
\equiv {\rho_{\oplus}^2 \over \rho_{\footnotesize\leftmoon}^2} 
 {R_{\oplus} \over R_{\footnotesize\leftmoon}} 
 { k_{2\footnotesize\leftmoon} \over Q_{\footnotesize\leftmoon} } {  Q_{\footnotesize\leftmoon} \over k_{2\oplus} }
  \approx 10 { k_{2\footnotesize\leftmoon}\over Q_{\footnotesize\leftmoon}} {Q_{\footnotesize\leftmoon}\over k_{2\oplus}} ,
\end{equation}

 that quantifies the relative importance of tidal dissipation in the Moon to tidal dissipation in the
 Earth\footnote{The power ``5'' is incorrectly written ``3'' in \citet{Touma1998} and \citet{Touma2000}.}
Today $k_{2{\footnotesize\leftmoon}}=0.02$ and $k_{2\oplus}=0.3$ \citep{Kaula1968},
which results in $A\approx 0.5$ (with $Q_{\oplus}=15$).
When $A\gg 1$, dissipation is relatively strong in the Moon, so that the orbit is circularized.
This is expected at early times when Earth is fully molten and $Q_{\oplus}$ is very high,
while the Moon has little or no atmosphere and thus quickly cools to a 
 highly dissipative state with $Q_{\footnotesize\leftmoon}$ in the range of 1 to 10.
On the other hand, when $A \ll 1$, dissipation in Earth is more important than dissipation in the Moon,
so that tides raise the eccentricity of the lunar orbit.
This is the case at later times when Earth's mantle is freezing and $Q_{\oplus}\ll 10$.

All models that we have looked at, with or without resonance capture
over a wide range of atmospheres and planetary albedos, predict (apart from an initial transient) the same evolution of $A$ as a function of $a$
(Figure \ref{fig:Figure10}; $A$ assumes current lunar parameters of
$k_{2\footnotesize\leftmoon}=0.02$ and $Q_{\footnotesize\leftmoon}=20$).  
The constancy of $A(a)$ (i.e., of $Q_{\oplus}(a)$, as $k_{2\oplus}$ does not stray from $1.5$ until late)
 implies that the tidal evolution of $Q_{\oplus}(a)$ is determined by a general principle, probably conservation of energy,
rather than by details of our model.

A second key point made in Figure \ref{fig:Figure10} is that, in all cases, $A\gg 1$ when the Moon encounters the evection
resonance. 
In the class of models being considered here,
there appears to be no way for tidal evolution to be both slow enough to hold the Moon within $a<7R_{\oplus}$ 
yet simultaneously to be highly dissipative (low $Q_{\oplus}$).
The reason for this is that small values of $Q_{\oplus}$ cause fast tidal evolution, quickly raising $a$ above $15R_{\oplus}$.  
Thus the Moon could not acquire much eccenticity while in the evection resonance.

A third point made in Figure \ref{fig:Figure10} is that, in all cases,
 $A \ll 1$ when $20R_{\oplus}\!<\!a\!<\!40 R_{\oplus}$.
 In this phase of lunar orbital evolution dissipation in the Earth would have been much more important than dissipation in the Moon,
so that tides raised in the Earth by the Moon would have raised the eccentricity of the lunar orbit.
How big $e$ can grow is unclear.
\citet{Touma1998} found that $e=0.5$ is achievable in the evection resonance. 
Presumably tidal flexure in the Moon caused by a highly eccentric
orbit would at some point trigger considerable tidal melting within the Moon and 
a consequent drop of $Q_{\footnotesize\leftmoon}\!$. 
Later still, after the Earth has mostly solidified, $A$ must rise again 
and most of any remaining eccentricity is taken out of the Moon's orbit.

How a late episode of lunar eccentricity might be twisted into today's lunar inclination is another puzzle
beyond the scope of this study.
\citet{Touma2000} mentions a secular resonance for growing inclination when the Moon was at $28R_{\oplus}$ if $e>0.6$,
but the mechanism is not otherwise described and we can say no more.
We do note that, in general, prospects for giving the Moon an inclination are probably best
when the Moon's orbit was neither wholly under the Sun's influence nor the Earth's influence,
which roughly corresponds to $15R_{\oplus}\!< \!a\!<\! 30R_{\oplus}$.  
During this interval the Laplace plane coincides with neither the ecliptic nor Earth's equator
which, if nothing else, creates an opportunity for helpful accidents \citep{Ward1981}.      
   
\section{Conclusions}

The greenhouse effect of a thick water-rich atmosphere slows the rate that a planet cools after a giant impact.
Here we show that for water-rich atmospheres,
the difference between the runaway greenhouse limit of $\sim 280$ W/m$^2$ and absorbed sunlight 
(which for the faint young Sun and and a planetary albedo of 0.3 is $\sim 170$ W/m$^2$) sets an upper bound of
$\sim 110$ W/m$^2$ on how quickly the Earth can cool. 
This in turn sets a $10^6$ year timescale for a magma surface
after an impact on the scale of the event that formed the Moon.

Tidal heating is a major term in the early Earth's energy budget. 
Tidal heating occurs mostly in mantle materials that are just beginning to freeze, which frustrates freezing.
The atmosphere's control over the planet's cooling rate 
sets up a negative feedback between viscosity-dependent tidal heating and the temperature-dependent viscosity of the magma ocean. 
While this feedback holds, the rate that the Moon's orbit evolves is limited by the modest radiative cooling rate of Earth's atmosphere,
which in effect tethers the Moon to the Earth. 
Consequently the Moon's orbit evolves orders of magnitude more slowly than in conventional models
that describe tidal evolution through fixed values of $Q$ \citep{Touma1998,Cuk2012}. 
The slow tidal evolution ensures that the evolving Earth-Moon system will be captured by
the strong evection resonance, but in our models the resonance encounter always occurs when tidal
dissipation is more efficient in the Moon than in the Earth, which suggests that capture by the evection resonance is unlikely to
generate much eccentricity in the lunar orbit. 
Whether significant angular momentum can be extracted from
the Earth-Moon system by cumulative solar torques from a low eccentricity lunar orbit is a question beyond the scope
of this research, but the answer to this question would have implications for \citet{Cuk2012}'s inventive hypothesis. 

Slow orbital evolution also promotes capture by other, weaker, higher order resonances \citep{Touma1998,Touma2000}.
Resonance capture by one or more of these resonances may have play a key role in the orbital evolution of the Earth-Moon system,
a possibility raised but not elaborated upon by \citet{Touma1998}.
Tidal dissipation in the Earth is most 
efficient at later times when the Earth's mantle has cooled enough to develop significant viscosity.
In all our models, tidal dissipation is most efficient in the Earth 
when the Moon is between $20R_{\oplus}$ and $40R_{\oplus}$.
During this interval it is likely that the Moon's eccentricity grew considerably.
The upper limit on eccentricity may be set by vigorous tidal heating in the Moon --- once the Moon melts,
tidal dissipation in the Earth and the Moon will be more comparable.
We speculate that this period of high eccentricity set the table for some other process,
perhaps capture by an otherwise unremarkable, little-noted resonance, that exploited its opportunity by leaving its mark on
the inclination of the Moon.   

\section*{Acknowledgments}
The authors thank the NASA Planetary Atmospheres Program for support of this work.

\end{document}